\documentclass[
 prl,
 reprint,
superscriptaddress,
 amsmath,amssymb,
 aps
]{revtex4-2}

\usepackage{physics}
\usepackage{graphicx}
\usepackage{dcolumn}
\usepackage{bm}
\usepackage{hyperref}
\usepackage{verbatim}    
\usepackage[dvipsnames]{xcolor}      
\usepackage[normalem]{ulem}   
\usepackage{placeins} 
\usepackage{makecell} 
\begin{document}

\preprint{APS/123-QED}

\title{Ultrafast light-induced long range antiferromagnetic correlations in paramagnets}

\author{Lorenzo Amato}
\affiliation{Condensed Matter Theory Group, Paul Scherrer Institute, CH-5232 Villigen PSI, Switzerland}
\affiliation{Laboratory for Solid State Physics, ETH Zürich, CH-8093 Zürich, Switzerland}
\author{Markus Müller}
\affiliation{Condensed Matter Theory Group, Paul Scherrer Institute, CH-5232 Villigen PSI, Switzerland}

\begin{abstract}
We propose and analyze a laser-driven protocol to generate long-range ordered patterns in paramagnets, based on 
non-adiabatically driven aggregation dynamics.
We derive the optimal driving parameters that maximize, respectively, the  correlation length or the size of defect-free antiferromagnetic clusters in a one-dimensional chain.
{We show that one can reach correlation lengths that are exponentially larger than those} achieved by adiabatic pumping schemes. The resonantly driven dynamics of cluster fronts is shown to map to an exactly solvable model of free fermions.
\end{abstract}

\maketitle
\paragraph{Introduction.}
The ultrafast creation of an ordered state of matter is a prime goal in solid state physics that will enable electronic or magnetic switches, as well as fast control of material properties. While certain orders are relatively straightforward to induce, e.g.  transient ferromagnetism created by the inverse Faraday effect using circularly polarized light \cite{kimel2005ultrafast}, spatially alternating patterns, such as transient  antiferromagnetism (AFM), are much harder to generate.

One way to induce such order is to transiently modify the effective parameters of a system by driving it. This is exploited, for instance, in Floquet engineering \cite{PhysRevB.103.L161106,PhysRevB.79.081406,moessner2017equilibration} and in adiabatic pumping schemes \cite{PhysRevLett.104.043002,keesling2019quantum} that slowly modulate drive frequency and amplitude. 
Alternative approaches avoid permanent driving, but rather excite the system into a metastable off-equilibrium state which develops order - e.g. through excitation-enhanced interactions \cite{wang2022light}, photodoping \cite{PhysRevB.102.165136,werner2019light} or by inducing structural transitions \cite{ichikawa2011transient,PhysRevB.84.241104}.
These methods rely on thermal relaxation within a kinetically restricted configuration space, {or under the action of a transiently modified Hamiltonian}. A more direct route, however, creates the desired order directly by non-adiabatic resonant aggregation dynamics \cite{garttner2013dynamic,valado2016experimental,lemeshko2012nonadiabatic}, building a pattern by tailored excitations that expand the domain walls surrounding an initially created seed. It was shown \cite{lemeshko2012nonadiabatic} that even with a single square pulse this scheme can achieve longer correlation lengths than adiabatic pumping. However, the great potential of using a sequence of optimally shaped driving pulses has  not been explored. In this Letter, we analytically study such a {\em multi-pulse resonant aggregation} and show that it can generate extremely large correlated AFM clusters.

\begin{figure}
    \centering
    \includegraphics[width=0.5\textwidth]{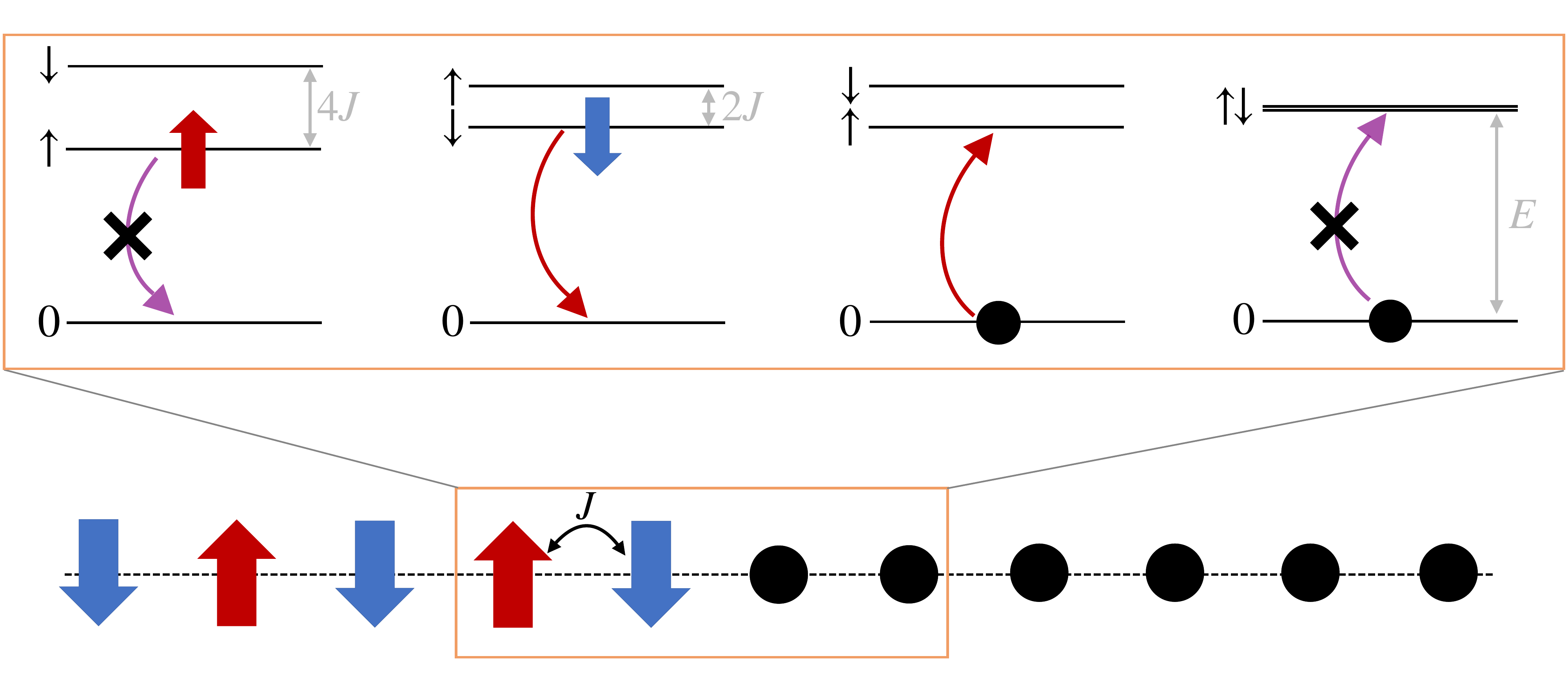}
    \caption{Level structure of ions around the right front of a growing AFM cluster in a chain governed by Eq.~\eqref{eq:H0}. The energy shifts due to the Ising interaction with excited neighbors allow ions at the front to be excited from the ground ($\ket{0}$) to the magnetized state ($\{\ket{\uparrow},\ket{\downarrow}\}$) matching the desired pattern.  
    Red arrows indicate resonant transitions for a driving frequency  $\omega=E-J$ (cf.~Eq.~\eqref{eq:HINT}). 
    The transitions marked by crossed purple arrows are off-resonant.}
    \label{fig:image1}
\end{figure}

\paragraph{Model.}
For simplicity we consider an insulating, one-dimensional chain of paramagnetic ions, each with a partially filled shell, as found in the rare earth (RE) family. We consider ions with an even number of electrons (non-Kramers), that have a non-magnetic singlet ground state $\ket{0}$, but possess a higher-lying, symmetry-protected crystal field doublet $\{\ket{\uparrow},\ket{\downarrow}\}$, which carries an Ising magnetic moment pointing along the crystalline $c$-axis.~\footnote{Note that excited quasi-doublets with splittings smaller than the Ising interaction are also acceptable.} 
Promising candidate materials are oxyborates (such as, e.g., Ca$_4$-RE-O(BO$_3$)$_3$~\cite{ExpMaterial} or certain perovskites.

If the excitation gap $E$ is either very small, or exceeds the phonon band width, such crystal field excitations can be very long lived \cite{orbach1961spin,malkin1987crystal}, as the phonon decay rate is very small.
We map the ground and the two excited doublet levels onto the states of an anisotropic, easy-plane $S=1$ pseudospin. Exchange or magnetic dipole coupling introduce  Ising interactions between the doublet states on neighboring ions, whereby we retain only the nearest neighbor (NN) interaction $J$, which for definiteness we assume  antiferromagnetic ($J>0$). An excitation on a given ion thus splits the doublet on its neighbors (cf.~Fig.~\ref{fig:image1}),
which results in the Hamiltonian
\begin{equation}
    H_0=\sum_{i=1}^{N}{E(S^{z}_{i})^2} +\sum_{<i,j>}{JS^{z}_{i}S^{z}_{j}}.
    \label{eq:H0}
\end{equation}
The crystal field splitting $E>0$  takes the role of a hard axis anisotropy, which we assume to be much larger than the interaction, $E\gg J$. {In general, there are also   magnetic interactions $J'$ that mediate flip-flop transitions between two ions that host a singlet and a doublet excitation, respectively. However, assuming a large magnetic moment of the doublet, the flip-flop interactions are significantly weaker than the longitudinal dipole interaction among doublet states, $J'\ll J$ \footnote{For specific point group symmetries (e.g. $D_6$) and certain irreducible representations thereof, one may have $J'=0$ due to selection rules}.} {Such additional interactions dress the classical Ising configurations by quantum fluctuations of the cluster edges, which renormalize the excitation energy and the Rabi frequency. However, the resonant aggregation dynamics is still well captured by an effective Hamiltonian as in (\ref{eq:H0}).
}

Spin lattice coupling introduces  a finite relaxation time $\tau_R$ for the excited ion states. Another important time scale is the decoherence time, $\tau_d \leq \tau_R$, over which the system can be considered as quantum coherent (while laser-driven). In practice, we expect $\tau_d$ to be limited by magnetic noise or phonon decay and to  substantially exceed the interaction time, $J\tau_d\gg 1$ \footnote
{We approximate $\tau_d$ as a parameter characterizing the material independently of the stepsize in the growth  of the driven AFM clusters. This  is justified if the dephasing due to dynamic fields in the material can be considered homogeneous over scales larger than a typical stepsize.}.

While our analysis focusses on a specific magnetic system, it can readily be adapted to other systems with similar interactions between neighboring sites, such as dipolar molecules \cite{lemeshko2012nonadiabatic} or Rydberg arrays \cite{garttner2013dynamic,valado2016experimental}.

\paragraph{Non-adiabatic resonant aggregation.} 
In pseudo-spin language, a laser drive takes the form of a time-dependent transverse field
\begin{equation}
    H_{\text{INT}}(t)=2\sqrt{2}\mathcal{E}(t)\cos{(\omega t)}\, \sum_{i=1}^{N}{S^{x}_{i}},
    \label{eq:HINT}
\end{equation}
where $\mathcal{E}(t)$ is proportional to the slowly varying amplitude envelope of the laser and the numerical prefactors have been chosen such that $\mathcal{E}(t)$ corresponds to the instantaneous (angular) Rabi frequency (cf. Eq.~\eqref{eq:HIINT} below).

\begin{figure}
    \centering
    \includegraphics[width=0.43\textwidth]{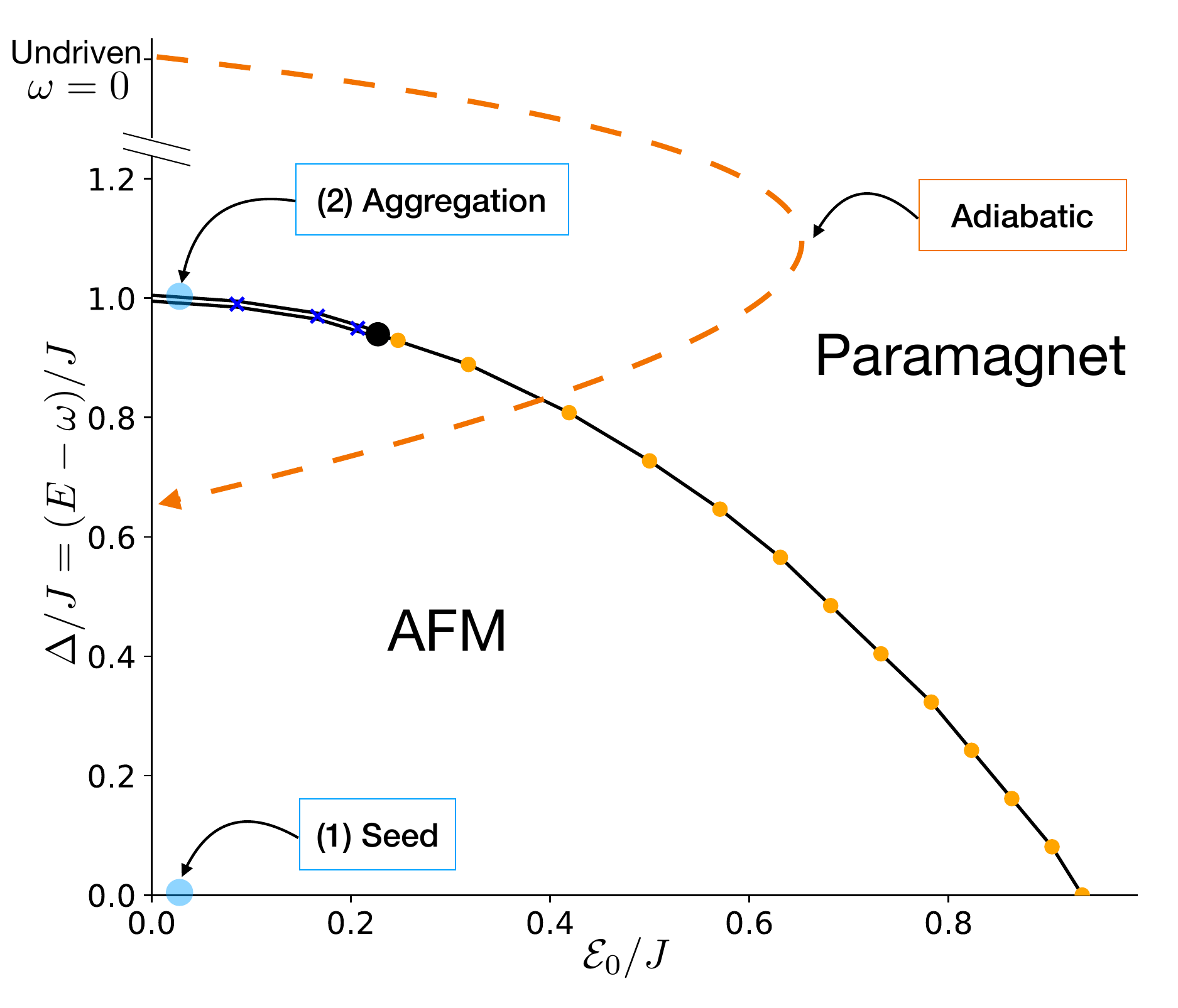}
    \caption{Phase diagram in the rotating frame as a function of the frequency detuning $\Delta=E-\omega$ and the amplitude ${\cal E}_0$. Blue crosses and yellow circles indicate the numerically determined location of discontinuous and continuous phase transitions, respectively, separated by a tricritical point (black dot). 
    An adiabatic single pulse protocol (orange arrow) crosses the continuous transition. However, the correlation length is bounded by the QKZ mechanism. Much larger correlation lengths are achieved in the non-adiabatic multi-pulse protocol: first AFM clusters are seeded with $\Delta =0$ and then grown by keeping parameters close to the border of the first order transition, $\Delta = J $ (light-blue dots).}
    \label{fig:phase_diag}
\end{figure}

A classical Néel pattern can be generated by driving with linearly polarized laser pulses in two steps, (cf.~Fig.~\ref{fig:phase_diag}):
First, a small density of doublet excitations is generated (for instance with a weak pulse of frequency $\omega = E$ [setting $\hbar=1$]). Then these seeds are made to expand by applying pulses whose frequency resonantly matches the energy $\omega_0\equiv E-J$ required to excite an ion  to the pattern-matching doublet state at the edges of existing clusters. As we will see, such a resonant driving induces ballistic propagation of the edges (cf.~Fig.~\ref{fig:image1}). 

The full time dependent problem $H(t)=H_0 + H_{\rm INT}(t)|_{\omega=\omega_0}$ is not solvable analytically. However, within a Rotating Wave Approximation (RWA) in the interaction picture, assuming a slow time dependence of the envelope $\mathcal{E}(t)$ and dropping terms oscillating with a high frequency of order $E$, we obtain
\begin{eqnarray}
H^I_{\text{RWA}}(t)=&&
\mathcal{E}(t)\sum_i\{e^{i[J+J(S^{z}_{i+1}+S^{z}_{i-1})]t}\dyad{i,\uparrow}{i,0} +\nonumber\\
&&+e^{i[J-J(S^{z}_{i+1} + S^{z}_{i-1})]t}\dyad{i,\downarrow}{i,0} + h.c.\}.
    \label{eq:HIINT}
\end{eqnarray}

The only time independent terms (i.e., resonant transitions) in Eq.~\eqref{eq:HIINT} are those of  spins at a cluster edge that have exactly one of their neighbors excited, and where the cluster grows or diminishes by one site, maintaining the AFM pattern. The non-resonant terms eventually introduce unintended defects whose probability we will estimate later.
Dropping them for the time being, we are left with the resonant approximation
\begin{eqnarray}
    H^I_{\text{R}}=&&
    \mathcal{E}(t)\sum_i\{\delta_{S^{z}_{i+1} + S^{z}_{i-1},-1}\dyad{i,\uparrow}{i,0} + \nonumber\\
    &&+\delta_{S^{z}_{i+1} + S^{z}_{i-1}, +1}\dyad{i,\downarrow}{i,0} + h.c.\},
    \label{eq:HIR}
\end{eqnarray}
in which the model becomes solvable \cite{ostmann2019localization,franchini2017introduction,majo_chain}: Under a Jordan-Wigner transformation,
{it maps to a tight-binding chain of fermions with dispersion $\epsilon_k(t)=-2\mathcal{E}(t)\cos{k}$. 
They describe the conserved edges  of AFM clusters},
which  move freely under the action of 
$H^I_{\text{R}}$. The fermionic exclusion principle reflects the fact that two cluster edges never occupy the same bond. {For the derivation and diagonalization of Eq.~\eqref{eq:HIR} we refer to the Supplemental Material (SM) \cite{SM}.} 
From the fermionic mapping it follows that under a resonant pulse, every cluster edge 
performs an independent Quantum Walk (QW), its wavefunction spreading over a range that increases linearly with time. Indeed, from the dispersion relation one can derive that the standard deviation of an edge's position evolves like $\sqrt{\ev{x^2}}=\sqrt{2}\mathcal{A}(t)$, where $\mathcal{A}(t)=\int_{-\infty}^t\mathcal{E}(t')\dd{t'}$ is the integrated pulse action.

\paragraph{Off-resonant transitions.}
Every driving protocol induces  
undesired transitions with finite probability. Errors come in three different types.  (i) In \emph{hole transitions} an ion inside a cluster flips back to the ground state $\ket{0}$. The transition energy is detuned from $\omega_0$ by $J$. Once a hole forms inside a cluster, it expands like the clusters do; nonetheless this will not destroy the correlation among the remaining excited spins. (ii) In a \emph{seed transition}, also off-resonant by $J$, a pseudo-spin is excited from $\ket{0}$ to a doublet state without having neighboring excited spins.
(iii) An \emph{edge transition} creates a misoriented excitation at the edge of a cluster, introducing a domain wall. 
This error type is suppressed as its energy mismatch is $2J$,
and will thus be neglected in the following. 

To minimize off-resonant transitions, we consider coherent Gaussian driving pulses with carrier frequency $\omega_0$, and an envelope with peak amplitude $\mathcal{E}_0$, temporal width $\sigma$ and sharp cut-offs at $t=\pm\tau/2$, 
 $\mathcal{E}(t)=\mathcal{E}_0 e^{-t^2/2\sigma^2}
  \Theta(\tau/2-|t|)$. 
Let us estimate the probability density $p_{\rm off}$  (per site in the bulk) to create a new seed, or to create a hole inside a cluster during a single pulse. We treat the resonant part to all orders \cite{SM} and include the off-resonant driving terms to leading order in time-dependent perturbation theory, which captures the scaling of $p_{\rm off}$. Up to numerical prefactors we find
\begin{eqnarray}
  \label{eq:p_off_app}
    p_{\rm off}(\mathcal{E}_0,\sigma,\tau) &&\approx
    \abs{\sum_{n\geq 0}\frac{\mathcal{F}\left[\mathcal{E}(t)\mathcal{A}^{2n}(t)\right](J)}{(2n)!}
    }^2 \\
    &&\approx \abs{\sum_{n\geq 0}\frac{1}{(J\sigma)^{2n}}\left[\hat{\mathcal{E}}\left(\frac{J}{2n+1}\right)\right]^{2n+1}
    }^2, \nonumber
\end{eqnarray}
{where $\mathcal{F}[\cdot]$ denotes Fourier transform}. $\hat{\mathcal{E}}(\omega)=\mathcal{F}[\mathcal{E}]$ is the spectrum of the pulse envelope, consisting of a Gaussian bulk  $\hat{\mathcal{E}}(\omega<\tau/2\sigma^2)\propto \mathcal{E}_0\sigma e^{-\omega^2\sigma^2/2}$ and a tail due to the cutoff $\hat{\mathcal{E}}(\omega>\tau/2\sigma^2)\propto \mathcal{E}_0 e^{-\tau^2/8\sigma^2}/\omega$.
{The second line in Eq.~\eqref{eq:p_off_app} follows from a saddle point approximation for large $J\sigma \gg 1$.}
The terms in Eq.~\eqref{eq:p_off_app} can be read as the amplitude of multi-photon processes, each involving the absorption/emission of a photon and the scattering of $n$ further photons, the dominant $(2n+1)$-photon processes involving frequencies $\omega\approx\omega_0 \pm J/(2n+1)$.
{As long as the pulse is not too wide ($\sigma<\sigma^*$), processes that scatter a typically large number $n^*(\mathcal{E}_0,\sigma)\approx\frac{J\sigma/2}{\sqrt{2\log{\frac{J}{\mathcal{E}_0}}+O(\log{\log{J\sigma}})}}$ of photons in the spectral bulk dominate, leading to $p_{\rm off}\sim \exp[- 2J \sigma \sqrt{2\log(J/\mathcal{E}_0)}]$.
For $\sigma>\sigma^*$, instead, the absorption/emission of a single photon from the tail with rate $p_{\rm off}\sim \hat{\mathcal{E}}^2(J)\sim \frac{\mathcal{E}_0^2}{J^2}e^{-\frac{\tau^2}{4\sigma^2}}$ is the leading process. The crossover occurs for a pulse width $\sigma^*\approx\left[\tau^2/(8J \sqrt{2\log{\frac{J}{\mathcal{E}_0}}})\right]^{1/3}$.}

\begin{figure}
    \centering
    \includegraphics[width=0.5\textwidth]{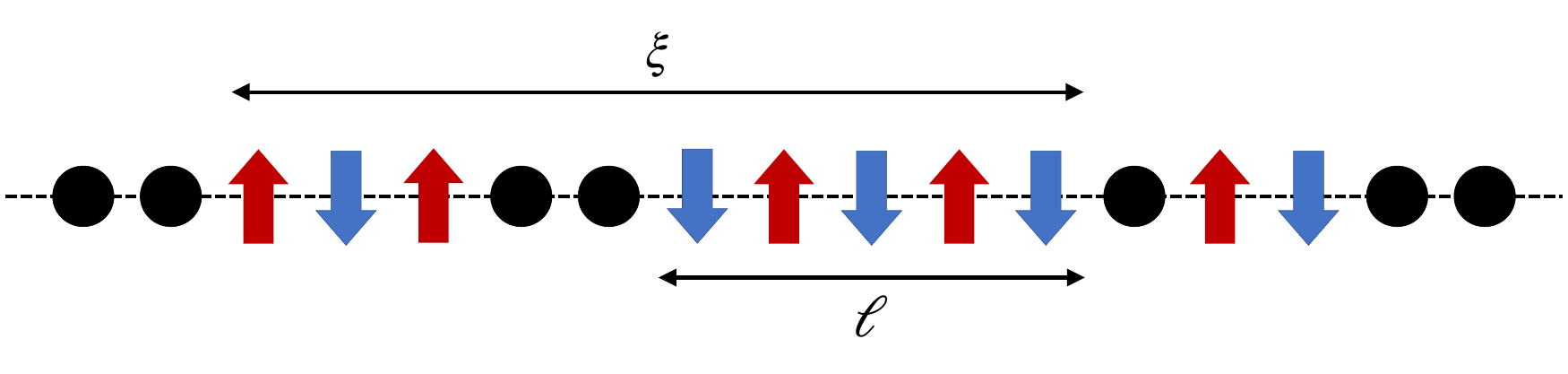}
    \caption{Visualization of the length $\ell$ of uninterrupted AFM clusters, and the usually much larger correlation length $\xi$, over which the {excited} ions maintain magnetic order.}
    
    \label{fig:chain_len}
\end{figure}

\paragraph{Optimal driving protocols.}
For a given interaction $J$ and times scales for decoherence ($\tau_d$) and relaxation ($\tau_R$), optimal drive parameters depend on the pursued goal. Typically, one either wants to maximize the AFM correlation length $\xi$ of excited regions, i.e., the range over which a given seed causes correlated excitations, ignoring potentially large holes within the emanating clusters; or, one aims for the largest possible defect-free clusters, whose size we denote by $\ell$, cf.~Fig.~\ref{fig:chain_len}.
When maximizing $\xi$, the expansion of correlations is ultimately limited by misaligned seeds created within the reach of the growing cluster. The spontaneous relaxation of excited spins does not affect the correlation among remaining excitations, but it certainly does limit the growth of hole-free clusters. In either case, the non-adiabatic driving profits from the ability of multiple pulses to create much bigger clusters than a single coherent pulse.

\paragraph{Optimizing the correlation length $\xi$.}
A single coherent pulse should be cut-off at $\tau_d$, as on longer time scales the dynamics  are incoherent. Hence we assume $\tau=\tau_d$. Under multiple pulses a cluster edge diffuses with diffusion constant $D=\mathcal{A}^2(\tau_d)/\tau_d \sim ({\cal E}_0\sigma)^2/\tau_d$ (setting the lattice constant to $a=1$). After $N_p$ consecutive pulses a typical cluster size scales as $\sim\sqrt{DN_p \tau_d}$. 

Consider the length scale on which the probability to generate at least one seed in an excited state uncorrelated with the growing cluster becomes $O(1)$. This will determine the correlation length $\xi$.
To reach this scale, one needs to apply $N_p\sim \xi^2/D\tau_d$. The probability of a ``seed error" within a range $\sim\xi$ occurring during any of these pulses scales as  $p_{\rm off} N_p \xi  \sim p_{\rm off} \xi^3 /D\tau_d$. Equating this to unity we infer $ \xi \sim \left({D\tau_d}/{p_{\rm off}}\right)^{1/3}$.

To maximize $\xi$  we should essentially minimize $p_{\rm off}(\mathcal{E}_0,\sigma,\tau_d)$ in  Eq.~\eqref{eq:p_off_app}. For a given drive amplitude $\mathcal{E}_0$, the optimal temporal pulse width is $\sigma_{\rm min}=\sigma^*$, which is much shorter than the cut-off time $\tau_d/2$. The maximal correlation length that can be obtained
then scales as
\begin{equation}
    \xi_{\rm max}(\mathcal{E}_0)\sim \exp{\frac{2^{1/3} }{3}(J\tau_d)^{2/3}\left(\log{\frac{J}{\mathcal{E}_0}}\right)^{1/3}},
\end{equation}
which requires a large number $N_p \sim \xi_{\rm max}^2(\mathcal{E}_0)$ of pulses.
Note that here we tacitly assume that the density of initial seeds does not exceed $1/\xi$. 
The above suggests that arbitrarily large $\xi$ can be reached by decreasing the pulse amplitude $\mathcal{E}_0$ (and thereby the error probability) while increasing the number $N_p$ of pulses. However, this overlooks the fact that $\mathcal{E}_0$ must be strong enough, such that the initial cluster seed is more likely to grow by at least one site, rather than to relax and disappear:  $(\mathcal{E}_0\sigma_{\rm min})^2>p_{\rm rel}
={\tau_d}/{\tau_R}$. This sets the lower bound $\mathcal{E}_0>\mathcal{E}_{0,c}\approx\left(\frac{J^2}{\tau_d\tau_R^3}\right)^{1/6}$, from which we derive the maximal correlation length of AFM-correlated regions,
\begin{equation}
\label{xiAFM}
    \xi_{\rm max}\sim \exp{\left(\frac{2}{3}\right)^{2/3}(J\tau_d)^{2/3}\left(\log\left[(J\tau_R)^3(J\tau_d)\right]
    \right)^{1/3}}.
\end{equation}
As we will derive below, the size of contiguous, hole-free AFM subclusters, is much smaller and scales as $\ell\sim \left(
D/p_{\rm rel}\right)^{1/3}\sim \left({\mathcal{E}_{0}}/{\mathcal{E}_{0,c}}\right)^{2/3}$. The last estimate holds {up to logarithmic corrections} for small drive amplitudes such that $p_{\rm off}< p_{\rm rel}$. The holes separating such contiguous subclusters can, however, have a diameter of order $O(\xi)$. 

The above result holds as long as the limiting seed errors result from multi-photon processes with  $n^*(\mathcal{E}_{0,c},\sigma_{\rm min})\gg 1$. This condition translates into $\log(J\tau_R)\ll J\tau_d$ which applies to most realistic situations. In the opposite case of extremely long-lived excitations the correlation length is limited by off-resonant transitions, which result in a maximal correlation length $\xi_{\text{AFM}}\sim \exp{\text{const.} \times J\tau_d}$ \cite{SM}. In this case the clusters are typically hole-free.

\paragraph{Optimizing the size of intact clusters $\ell$.}
The size of defect-free clusters $\ell$ is limited by unintended hole transitions where an ion relaxes back to its ground state. We thus have $\ell\sim\left(
{D}/{p_{\rm hole}}\right)^{1/3}$. Note that the hole creation probability (per site and per pulse) receives  contributions both from driving and spontaneous relaxation, 
$p_{\rm hole}=p_{\rm off}+p_{\rm rel}$.
The driven hole transitions are much less frequent than spontaneous relaxation, unless the temporal pulse width {$\sigma$ approaches the cut-off $\tau_d$ which results in a fat  tail of the pulse spectrum}. Since $D\sim \sigma^2$, it thus pays to increase $\sigma$ (beyond $\sigma_{\rm min}$) until $p_{\rm off}\sim p_{\rm rel}$, which yields $\sigma_{\rm opt}\approx \tau_d/2\sqrt{\log[(\mathcal{E}_0/J)^2/p_{\rm rel}]}$.
This is optimal since beyond that crossover $p_{\rm hole }\sim p_{\rm off}$ increases faster than $D$ with increasing $\sigma$.

The diffusion constant $D$ is maximized by a large Rabi frequency $\mathcal{E}_0\to \mathcal{E}_{\rm max}$. If it is not limited by laser power one may choose $\mathcal{E}_{\rm max}$ of the order of, but still logarithmically smaller than $J$ (to maintain a clear distinction between resonant and off-resonant transitions).
The largest contiguous clusters thus reach a size of $\ell_{\rm max}\sim [(\sigma_{\rm opt} \mathcal{E}_{\rm max})^2/p_{\rm rel}]^{1/3} \sim \left(
\mathcal{E}_{\rm max}^2\tau_d\tau_R\right)^{1/3}
$ - up to logarithmic corrections.
They are generated by a sequence of $N_p\sim\frac{\ell_{\rm max}^2}{D\tau_d}\sim\left(\frac{\tau_R}{\tau_d^2 \mathcal{E}_{\rm max}}\right)^{2/3}$
strong and long pulses. In the above we have assumed that the relaxation time was sufficiently long ($\tau_R > \mathcal{E}_{\rm max} \tau_d^2$, and thus $N_p>1$). 
If instead $\tau_R$ is shorter, the optimal protocol consists in a single pulse, the size of contiguous clusters being limited by the condition $\ell\, p_{\rm rel}< 1$, implying $\ell_{\rm max}\sim
{\tau_R}/{\tau_d}$.

\paragraph{Comparison with adiabatic driving.}
It is interesting to compare resonant aggregation with an adiabatic driving protocol, which attempts to follow the ground state of the RWA Hamiltonian in the rotating frame,
\begin{equation}
 H_{\text{RWA}}(\Delta,\mathcal{E})=\sum_{i=1}^N\Delta (S^z_n)^2 + \sum_{<i,j>}{JS^{z}_{i}S^{z}_{j}} + \sum_{i=1}^N\sqrt{2}\mathcal{E}S^x_n,
\end{equation}
as the detuning $\Delta= E-\omega$, and the amplitude $\mathcal{E}$ of the coherent pulse are tuned across the quantum phase transition (QPT) to the AFM ordered phase. 
Fig.~\ref{fig:phase_diag} shows the transition line between paramagnet and AFM as obtained via finite size scaling analysis based on the Binder cumulant \cite{binder1981static}, cf.~the SM \cite{SM}.
A multi-critical point separates a first order transition (between the classical limit $\mathcal{E}=0$ and $\mathcal{E}< 0.23 J$) from a continuous transition in the Ising universality class, as is expected from the equivalence of this $1D$ quantum model with the $2D$ classical Blume-Capel model \cite{blume_orig,capel1966possibility,zierenberg2017scaling,sachdev2011quantum}.

The crossing of the transition necessarily becomes non-adiabatic at some point and the achievable correlation length is limited by the unavoidable generation of defects via the quantum Kibble-Zurek (QKZ) mechanism \cite{zurek2005dynamics,keesling2019quantum}. Note that the adiabatic preparation relies on coherence and  is thus limited to a single pulse of duration $\sim \tau_d$. In contrast, aggregation dynamics starts from dilute excited nucleation centers, and repeated pulses allow the exploration of configurations resonant with the initial state; hence, the only limitations are undesired off-resonant transitions and relaxations to non-targeted parts of phase space, cf.~Fig.~\ref{fig:phase_diag}.
For the quasi-adiabatic scheme, the QKZ mechanism predicts a correlation length $\xi_{\rm QKZ}\sim (J\tau_d)^{1/2}$ for a sweep at constant rate across the Ising transition.~\cite{zurek2005dynamics,cui2020experimentally} 
This is shorter than $\xi_{\rm sq}\sim \left(J\tau_d\right)^{2/3}$ which non-adiabatic aggregation dynamics achieves
with a single square pulse~\cite{lemeshko2012nonadiabatic}.
{By adapting the rate in the  adiabatic passage through the phase transition~\cite{PhysRevLett.101.076801} one can improve to $\xi_{\rm QKZ, opt}\sim J\tau_d/\log{J\tau_d}$, which is, however, still slightly smaller than $\xi_{\rm opt}\sim J\tau_d/\sqrt{\log{J\tau_d}}$, obtained with an optimal non-adiabatic Gaussian drive~\footnote{For a single pulse the coherent cluster size $\xi$ is limited by $\xi\lesssim 1/p_{\rm off}$. For a square pulse,  $p_{\rm off}=\frac{\mathcal{E}_0^2}{J^2}$ if $\mathcal{E}_0\ll J$, while the cluster grows to $\xi\sim \mathcal{E}_0\tau_d$. The  amplitude maximizing $\xi$ is thus  $\mathcal{E}_0\sim\left(\frac{J^2}{\tau_d}\right)^{1/3}$,  resulting in $\xi_{\rm sq}\sim \left(J\tau_d\right)^{2/3}$~\cite{lemeshko2012nonadiabatic}. Similarly, for optimal Gaussian pulses, using $p_{\rm off}\approx \frac{\mathcal{E}_0^2}{J^2}\exp{-\frac{\tau_d^2}{4\sigma^2}}$ and $\xi\sim \mathcal{E}_0\sigma$. The optimal driving parameters are $\mathcal{E}_0\lesssim J$ and $\sigma \approx \frac{\tau_d}{2(\log{J\tau_d})^{1/2}}$, up to logarithmic corrections.}}, but exponentially smaller than the multi-pulse result \eqref{xiAFM}.

\paragraph{Conclusions.} We have shown that a non-adiabatic resonant aggregation scheme in quasi-1d non-magnetic insulators can induce AFM correlations up to scales that are exponentially large in the parameter $(J\tau_d)^{2/3}$, much larger than those within   quasi-adiabatic reach. This requires driving with multiple Gaussian pulses, that minimize the creation of order-breaking excitations. The very long correlations come at the price of relatively large intervals of un-excited  ions within one correlated region.
If instead the aim is to maximize uninterrupted AFM clusters, one should apply fewer long and strong pulses. Also this results in clusters  parametrically larger than those obtained with adiabatic driving. 

While the model discussed here considers nearest-neighbor interactions only, the scheme can be extended to longer range (e.g., dipole) interactions, where more complex, chirped pulses might be required in the initial stages of the cluster growth.

It will be interesting to extend the analysis of optimal aggregation dynamics to higher dimensions. A simple translation of our proposed driving scheme to bipartite lattices is expected to result in the growth of self-avoiding branched polymer structures with AFM correlations extending over a diameter comparable to the one obtained here for the 1d chain.  
On top of that, given the possibility of spontaneous symmetry breaking and genuine long-range order of Ising models in $d>1$, it will be interesting to study whether restricted thermalization (conserving the number of excited ions, and possibly of domain walls) after the laser-driven excitation  can substantially extend the spatial correlations by coarsening dynamics.

{\paragraph{Acknowledgments.}
We thank G. Aeppli, M. Kenzelmann, G. Matmon and S. Nikitin for discussions. We acknowledge financial support from the Swiss National Science Foundation under grant No. 200020$\_$200558.
}
\bibliography{biblio}
\end{document}


\preprint{APS/123-QED}

\title{Supplemental Material: Ultrafast light-induced long range antiferromagnetic correlations in paramagnets}

\author{Lorenzo Amato}
\affiliation{Condensed Matter Theory Group, Paul Scherrer Institute, CH-5232 Villigen PSI, Switzerland}
\affiliation{Laboratory for Solid State Physics, ETH Zürich, CH-8093 Zürich, Switzerland}
\author{Markus Müller}
\affiliation{Condensed Matter Theory Group, Paul Scherrer Institute, CH-5232 Villigen PSI, Switzerland}

\maketitle
\section{\label{sec:supp1}
Rotating Wave Approximation of the driven Hamiltonian in the interaction picture}
In this section, we derive Eq.~(3) of the main text from the Hamiltonian $H_{\text{INT}}$ (see Eq.~(2) there) in the interaction picture $H^I_{\text{INT}}=e^{iH_0t}H_{\text{INT}}e^{-iH_0t}$, with $H_0$ defined in Eq.~(1) of the main text. $H^I_{\text{INT}}$ can be expressed in simpler terms if we use the formula:
\begin{equation}
    e^{A}Be^{-A}=B + [A,B] + \frac{1}{2!}[A,[A,B]]+...
\end{equation}
and decompose $S^{x}$ (written here in the $S^z$ basis) as
\begin{eqnarray}
&&S^{x}=\frac{1}{\sqrt{2}}\begin{pmatrix}
0 && 1 && 0\\
1 && 0 && 1\\
0 && 1 && 0\\
\end{pmatrix}=\frac{1}{\sqrt{2}}\left[\begin{pmatrix}
0 && 1 && 0\\
0 && 0 && 0\\
0 && 0 && 0\\
\end{pmatrix} +\right.\nonumber\\ 
+&&\begin{pmatrix}
0 && 0 && 0\\
1 && 0 && 0\\
0 && 0 && 0\\
\end{pmatrix} + \begin{pmatrix}
0 && 0 && 0\\
0 && 0 && 1\\
0 && 0 && 0\\
\end{pmatrix} + \left.\begin{pmatrix}
0 && 0 && 0\\
0 && 0 && 0\\
0 && 1 && 0\\
\end{pmatrix}  \right]=\nonumber\\
\equiv && S^{+,+} + S^{-,-} + S^{-,+} + S^{+,-}.
\end{eqnarray}
We consider a 1D chain with $N+2$ sites, labelled by $i\in\{0,\dots, N+1\}$. We assume $N$ to be odd for simplicity. In the following, we will be interested in the bulk of the chain consisting of the sites $\{1,\dots,N\}$. It is easy to show that for $S^{\lambda,\lambda'}_i$:
\begin{equation}
    [H_0,S_i^{\lambda,\lambda'}]=[\lambda E +\lambda' J(S^{z}_{i+1} + S^{z}_{i-1})]S^{\lambda,\lambda'}_i\, ,
\end{equation}
where $\lambda=\pm$ and $\lambda'=\pm$, and $i$ is in the bulk. This implies that
\begin{equation}
     \underbrace{[H_0,[H_0,...[H_0}_{n \text{ nestings}},S^{\lambda,\lambda'}_i]]]=[\lambda E +\lambda' J(S^{z}_{i+1} + S^{z}_{i-1})]^{n}S^{\lambda,\lambda'}_i.
\end{equation}
It is then possible to rewrite the bulk terms of $H_{\text{INT}}^I$ as
\begin{eqnarray}
H^I_{\text{INT}}(t)=&&2\sqrt{2}\mathcal{E}(t)\cos{(\omega t)}\sum_{i=1}^{N}\{e^{i[E+J(S^{z}_{i+1} + S^{z}_{i-1})]t}S^{+,+}_i +\nonumber\\
&&+e^{i[E-J(S^{z}_{i+1} + S^{z}_{i-1})]t}S^{+,-}_i + h.c.\}.
    \label{eq:pre_RWA}
\end{eqnarray}
With the frequency $\omega$ of the laser pulse set to resonantly match the transition at $\omega=E-J$, we proceed with the Rotating Wave Approximation (RWA), removing terms with high frequencies of order $E$, retaining only terms oscillating with frequencies of order $J (\ll E)$:
\begin{eqnarray}
H^I_{\text{RWA}}(t)=&&\sqrt{2}\mathcal{E}(t)\sum_{i=1}^N\{e^{i[J+J(S^{z}_{i+1} + S^{z}_{i-1})]t}S^{+,+}_i + \nonumber\\
+&&e^{i[J-J(S^{z}_{i+1} + S^{z}_{i-1})]t}S^{+,-}_i + h.c.\}
    \label{eq:H_RWA}
\end{eqnarray}
which is equivalent to Eq.~(3) of the main text, noting that:
\begin{eqnarray}
    \sqrt{2}S^{+,+}_i=\dyad{i,\uparrow}{i,0};\nonumber \\
    \sqrt{2}S^{+,-}_i=\dyad{i,\downarrow}{i,0};\nonumber \\
    \sqrt{2}S^{-,-}_i=\dyad{i,0}{i,\uparrow};\nonumber \\
    \sqrt{2}S^{-,+}_i=\dyad{i,0}{i,\downarrow}.
\end{eqnarray}

\section{Mapping to a Majorana chain}
\label{sec:supp2}
Here we formally derive the mapping from the Hamiltonian of Eq.~(4) of the main text to a solvable model of a Majorana chain. Suppose we initially seed a spin $\ket{\uparrow}$ excitation on site $i$. Subsequent resonant transitions can only flip spins only to the $\ket{\downarrow}$ state on sites at an odd distance to $i$, and only to $\ket{\uparrow}$ on sites at an even distance. This suggests the possibility of reducing the dimensionality of our Hilbert space by mapping our model to an effective spin-$\frac{1}{2}$ model: for a given site, the two states are $\ket{0}$ and the resonantly reachable excited state. One can easily check that the resonant Hamiltonian $H^I_{\text{R}}$ of Eq.~(4) in the main text effectively maps to
\begin{eqnarray}
    &&H_{1/2}=\mathcal{E}(t)\sum_{j=1}^N\delta_{Z_{j-1} +Z_{j+1},0}X_j=\nonumber\\
    &&=\frac{\mathcal{E}(t)}{2}\left(\sum_{j=1}^NX_j - \sum_{j=1}^{N}Z_{j-1}X_{j}Z_{j+1}\right),
\end{eqnarray}
where the Pauli matrix $Z_i$ acts diagonally in the two relevant basis states of site $i$, while $X_i$ swaps them.

We employ a Jordan-Wigner transformation {anchored at the boundary site $0$} \cite{jordan1993paulische,majo_chain}
\begin{eqnarray}
    \xi_j=X_0\dots X_{j-1}Y_j,\nonumber\\
    \eta_j=X_0\dots X_{j-1}Z_{j},
    \label{eq:majo}
\end{eqnarray}
where $0\leq j \leq N+1$ (with $\xi_0=Y_0$ and $\eta_{0}=Z_0$). The operators $\xi_j=\xi_j^\dagger$ and $\eta_j= \eta_j^\dagger$ are Majorana fermions, obeying anticommutation relations:
\begin{eqnarray}
    \{\xi_i,\xi_j\}&&=\{\eta_i,\eta_j\}=2\delta_{i,j},\\
    \{\xi_i,\eta_j\}&&=0.
\label{eq:clif}
\end{eqnarray}
The Hamiltonian now reads
\begin{equation}
    H_{1/2}=-\frac{\mathcal{E}(t)}{2}\sum_{j=1}^{N}i\left(\xi_j\eta_{j}{+}\xi_{j-1}\eta_{j+1}\right),
\end{equation}
which can be mapped to the Hamiltonian of a nearest-neighbor tight-binding chain
\begin{eqnarray}
    H_{1/2}=&&-\mathcal{E}(t)\sum_{j=1}^N(c^\dagger_jc_{j+1} + h.c.)=\nonumber\\
    =&&-2\mathcal{E}(t)\sum_{k\in BZ}\tilde{c}^{\dagger}_k\tilde{c}_k\cos{k},
\end{eqnarray}
by introducing the following fermionic operators (for $1\leq l \leq (N+1)/2$):
\begin{eqnarray}
    c_{2l-1}=\frac{1}{2}(\eta_{2l} +i\eta_{2l-1}),\nonumber\\
    c_{2l}=\frac{1}{2}(\xi_{2l-1} -i\xi_{2l}),
\label{eq:F_JW}
\end{eqnarray}
and the Fourier transform
\begin{equation}
    \tilde{c}_k=\frac{1}{\sqrt{N}}\sum_{j=1}^Ne^{-ikj}c_j.
\end{equation}
The time evolution of the fermionic operator $c_j$ under the Hamiltonian $H_{1/2}$ is
\begin{eqnarray}
    &&c_j(t)=\frac{1}{\sqrt{N}}\sum_{k\in BZ}e^{ikj}e^{2i\mathcal{A}(t)\cos{k}}\tilde{c}_k=\nonumber\\
    &&=\sum_{n=1}^{N}i^{j-n}J_{j-n}(2\mathcal{A}(t))c_n,
\end{eqnarray}
where BZ denotes the Brillouin zone and 
\begin{eqnarray}
\mathcal{A}(t)=\int_{-\infty}^{t}\mathcal{E}(t')\dd{t'},
\end{eqnarray}
 and $J_{\alpha}(x)$ is the Bessel function of the first kind of order $\alpha$.

We now introduce a new species of fermionic domain wall operators $\gamma_j$:
\begin{eqnarray}
    \gamma_{2m-1}&=&\eta_{2m},\nonumber\\
    \gamma_{2m}&=&i\xi_{2m}.
\end{eqnarray}
We will refer to these operators as ``Majoranas" even though $\gamma_j^\dagger=(-1)^{j+1}\gamma_j$. These operators anticommute and $\gamma_j\gamma_j^\dagger=1$. $\gamma_j$ is  associated with the creation or annihilation of a domain wall on the bond between  sites $j$ and $j+1$, hosting a $\ket{\uparrow}$ (or $\ket{\downarrow}$) and a $\ket{0}$ state, respectively, and flipping the excitation state of all spins to the left of that bond. 
Note that we only used the even-labelled Majoranas $\eta$ and $\xi$, since the odd-labelled Majoranas act very similarly.
These domain wall Majoranas evolve as
\begin{equation}
   \gamma_{m}(t)=\sum_{n=1}^{N}(-i)^{m-n}J_{m-n}(2\mathcal{A}(t))\gamma_{n}.
    \label{eq:gamma_t}
\end{equation}

The above formalism allows us to calculate the transition probability between different domain wall configurations in a convenient way. 
 Any product state configuration of domain walls can be written as the result of acting with multiple $\gamma_j$ operators on the $\ket{\bm{0}}$ state. We adopt the notation $\ket{\bm{a}}$, with $\bm{a}=\{a_1,a_2,\dots,a_M\}$ being an ordered non-repeating $M$-tuple, to indicate a configuration with $M$ domain walls on bonds $(a_i, a_i+1)$, and explicitly express it as
\begin{equation}
    \ket{\bm{a}}=\gamma^\dagger_{a_1}\dots\gamma^\dagger_{a_M}\ket{\bm{0}}\, .
    \label{eq:multi_gamma}
\end{equation}
We can now calculate the transition amplitude between any domain wall configuration as long as the number of domain walls is conserved. The latter is indeed the case for resonant transitions. Let us consider the two configurations $\ket{\bm{a}}$ and $\ket{\bm{a'}}$, each containing $M$ domain walls. The transition probability between these two states is
\begin{equation}
    P_{\bm{a}\to\bm{a'}}(t)=\abs{\ev{\gamma_{a'_1}\dots\gamma_{a'_{M}}\gamma^\dagger_{a_1}(t)\dots\gamma^\dagger_{a_{M}}(t)}_{\rm vac}}^2,
\end{equation}
where $\ev{\dots}_{\rm vac}$ is the expectation value over the $\ket{\bm{0}}$ state. This can be recast, by making the time dependence explicit upon using Eq~\eqref{eq:gamma_t}, into
\begin{eqnarray}
&&P_{\bm{a}\to\bm{a'}}=\left|\sum_{\bm{\beta}}\ev{\gamma_{a'_1}\dots\gamma_{a'_{M}}\gamma^\dagger_{\beta_1}\dots\gamma^\dagger_{\beta_{M}}}_{vac}\times\right.\nonumber\\
&&\left.\times\prod_{n=1}^{M}J_{a_n-\beta_n}(2\mathcal{A}(t))i^{a_n-\beta_n}\right|^2.
\label{eq:trans_majo}
\end{eqnarray}
where $\sum_{\bm{\beta}}=\sum_{\beta_1}\dots\sum_{\beta_{M}}$. 
 By using the properties of the $\gamma$ operators, one can see that the only non-zero terms of the sum are those for which the $M$-tuple $\bm{\beta}$ is a permutation of the $M$-tuple $\bm{a'}$. Each index swap in the $M$-tuple $\bm{\beta}$ comes with a minus sign, and their totality can be factored into the sign of the permutation. A global phase  $\exp{i\frac{\pi}{2}\left(\sum_{n=1}^Ma_n -\sum_{n=1}^Ma'_n\right)}$ can be factored out of the sum. We then finally obtain the transition probability
\begin{equation}
P_{\bm{a}\to\bm{a'}}(t)=\left(\det{J^{\bm{a}\to\bm{a'}}}\right)^2,
\label{eq:prob_trans_majo}
\end{equation}
where the matrix $J^{\bm{a}\to\bm{a'}}_{i,j}=J_{a_i-a'_j}(2\mathcal{A}(t))$.

\section{Off-resonant transition probabilities}
In this section we perform a perturbative calculation of the probabilities, $p_{\rm hole}$ and $p_{\rm seed}$, {for off-resonant hole and seed transitions, respectively}. We first derive and discuss $p_{\rm seed}$. In the end, we argue that $p_{\rm hole}$ follows the same parametric dependence, which justifies why we refer to both probabilities summarily as $p_{\rm off}$ in the main text.

Let us derive the result given in Eq.~(5) {and later parts of the main text}. The creation of a new seed on site $i$ during a pulse corresponds to the creation of two neighboring domain walls, $\gamma_{i-1}\gamma_{i}$, which subsequently can propagate resonantly. 

We may express the leading seed-creating off-resonant terms in Eq.~\eqref{eq:H_RWA} in terms of domain wall operators.  
A representative term takes the form
\begin{equation}
    H_{\rm seed}(t)=\mathcal{E}(t)\sum_n \left(e^{iJt}\gamma^\dagger_{n-1}\gamma^\dagger_n+h.c.\right),
\end{equation}
from which we estimate the transition probabilities. The above expression omits  projection factors that make sure that $H_{\rm seed}$ always changes the number of domain walls by $\pm 2$. Those matter when domain walls are present in the vicinity of where the seed is to be created. {However, these projectors are irrelevant when we describe the generation of a  single seed by acting with  $H_{\rm seed}$ on the ground state.}

The simplest way to compute $p_{\rm seed}$ is to calculate the transition amplitude $\alpha_{i,j}$ for the process $\ket{\bm{0}}\to\gamma_{i}\gamma_j\ket{\bm{0}}$ in time-dependent perturbation theory in the interaction picture. For an order of magnitude estimate we may restrict ourselves to the first order in $H_{\rm seed}$ as it is the leading term that relates states with the smallest energy mismatch $J$. Terms of higher order in $H_{\rm seed}$ are expected to be smaller, especially when typical domain walls separations are $\gg 1$. 
To first order we find
\begin{equation}
    \alpha_{i,j}=-i\sum_n\int_{-\infty}^{\infty}{\mathcal{E}(t)e^{iJt}\ev{\gamma_i(t)\gamma_j(t)\gamma^\dagger_{n-1}\gamma^\dagger_n}{\bm{0}}\dd{t} },
\end{equation}
which, upon using Eq.~\eqref{eq:trans_majo} equals
\begin{eqnarray}
    &&\alpha_{i,j}=e^{i\frac{\pi}{2}(i+j)}\sum_n(-1)^n\int_{-\infty}^{\infty}\mathcal{E}(t)e^{iJt}\left[J_{i-n+1}(2\mathcal{A}(t))\times\right.\nonumber \\
    &&\left.\times J_{j-n}(2\mathcal{A}(t))-J_{i-n}(2\mathcal{A}(t))J_{j-n+1}(2\mathcal{A}(t))\right]\dd{t}.
\end{eqnarray}
The summation over the index $n$ can be simplified by using the summation theorem for Bessel functions (assuming that $n$ runs over all of $\mathbb{Z}$, which is a good approximation inside the bulk) \cite{gradshteyn2014table}
\begin{equation}
    \sum_{r\in\mathbb{Z}}(-1)^rJ_r(x)J_{r+\nu}(x)=J_{\nu}(2x).
\end{equation}
In our case we use it for the first term with $r\to i-n+1$ and $\nu\to -i+j-1$, and for the second term with $r\to i-n$ and $\nu\to -i+j+1$. This yields
\begin{eqnarray}
    \alpha_{i,j}=&&-e^{i\frac{\pi}{2}(i-j)}\int_{-\infty}^{\infty}\mathcal{E}(t)e^{iJt}\left[J_{j-i-1}(4\mathcal{A}(t))+\right.\nonumber\\
   &&\left.\quad\quad  +J_{j-i+1}(4\mathcal{A}(t))\right]\dd{t}.
\end{eqnarray}
This can be further simplified using the recurrence relations of Bessel functions as
\begin{eqnarray}
    \alpha_{i,j}=&&-e^{i\frac{\pi}{2}(i-j)}\int_{-\infty}^{\infty}\mathcal{E}(t)e^{iJt}\frac{j-i}{2\mathcal{A}(t)}J_{j-i}(4\mathcal{A}(t))\dd{t}.\nonumber\\
\end{eqnarray}
Due to translation symmetry, $\alpha_{i,j}$ is only a function of $i-j$ (which becomes exact for  $i$ and $j$ far from the sample boundaries).

The probability density (per site in the bulk) of creating a seed during a pulse is then
\begin{eqnarray}
\label{eq:p_seed_fin}
    p_{\rm seed}=&&\sum_{j\in\mathbb{Z}}\abs{\alpha_{i,i+j}}^2\\
    =&& \sum_{j\in\mathbb{Z}}\frac{j^2}{4}\biggl|\int_{-\infty}^{\infty}\mathcal{E}(t)e^{iJt}\frac{J_{j}(4\mathcal{A}(t))}{\mathcal{A}(t)}\dd{t}\biggr|^2.\nonumber
\end{eqnarray}
We focus on the case of a Gaussian pulse of total duration $\tau$ and having a characteristic pulse width $\sigma$, 
\begin{equation}
\mathcal{E}(t)=\mathcal{E}_0 e^{-t^2/2\sigma^2}\Theta(\tau/2-|t|).
\end{equation}  

\begin{figure}[ht!]
    \centering
    \includegraphics[width=0.48\textwidth]
    {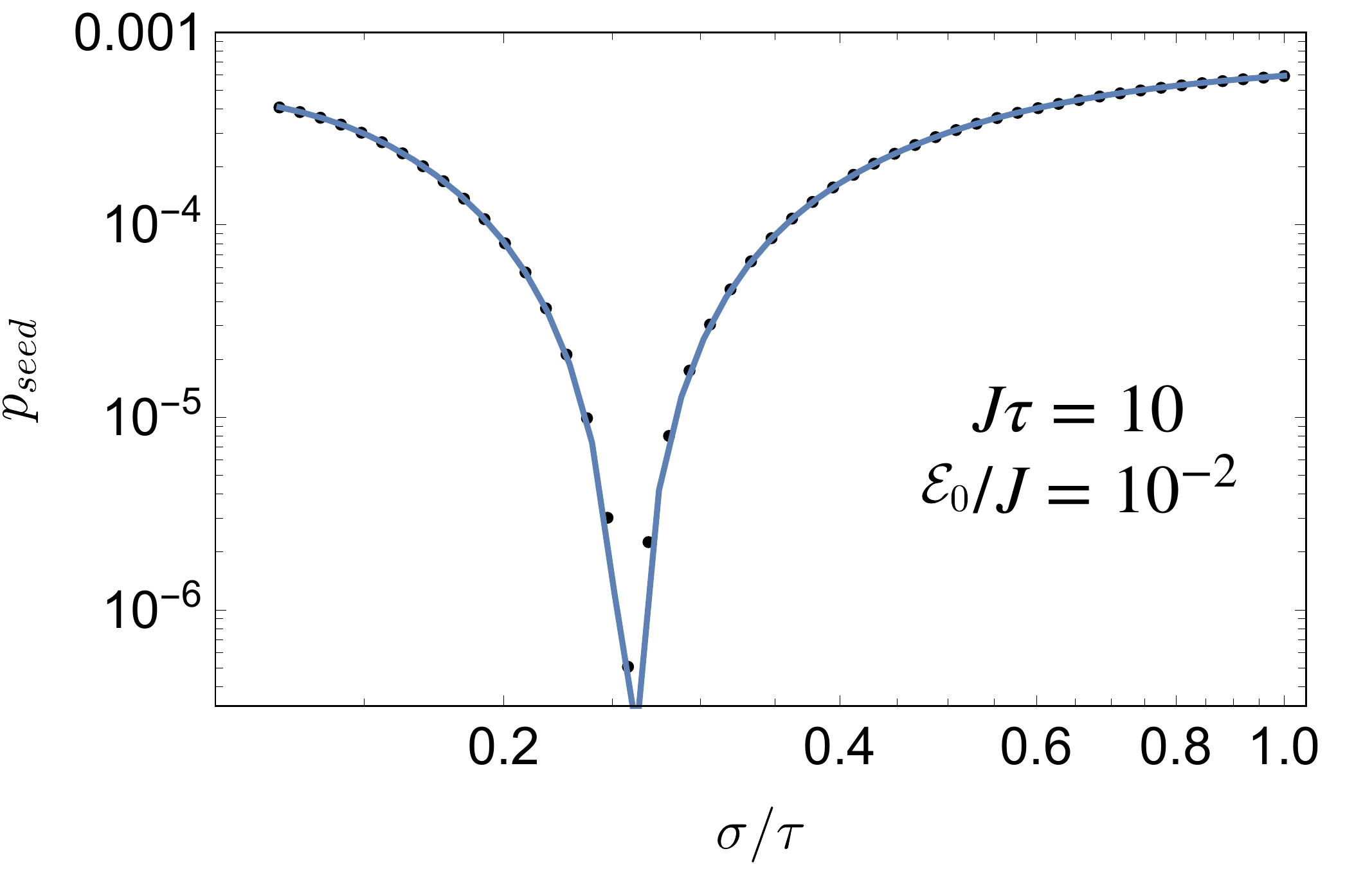}
    \includegraphics[width=0.48\textwidth]{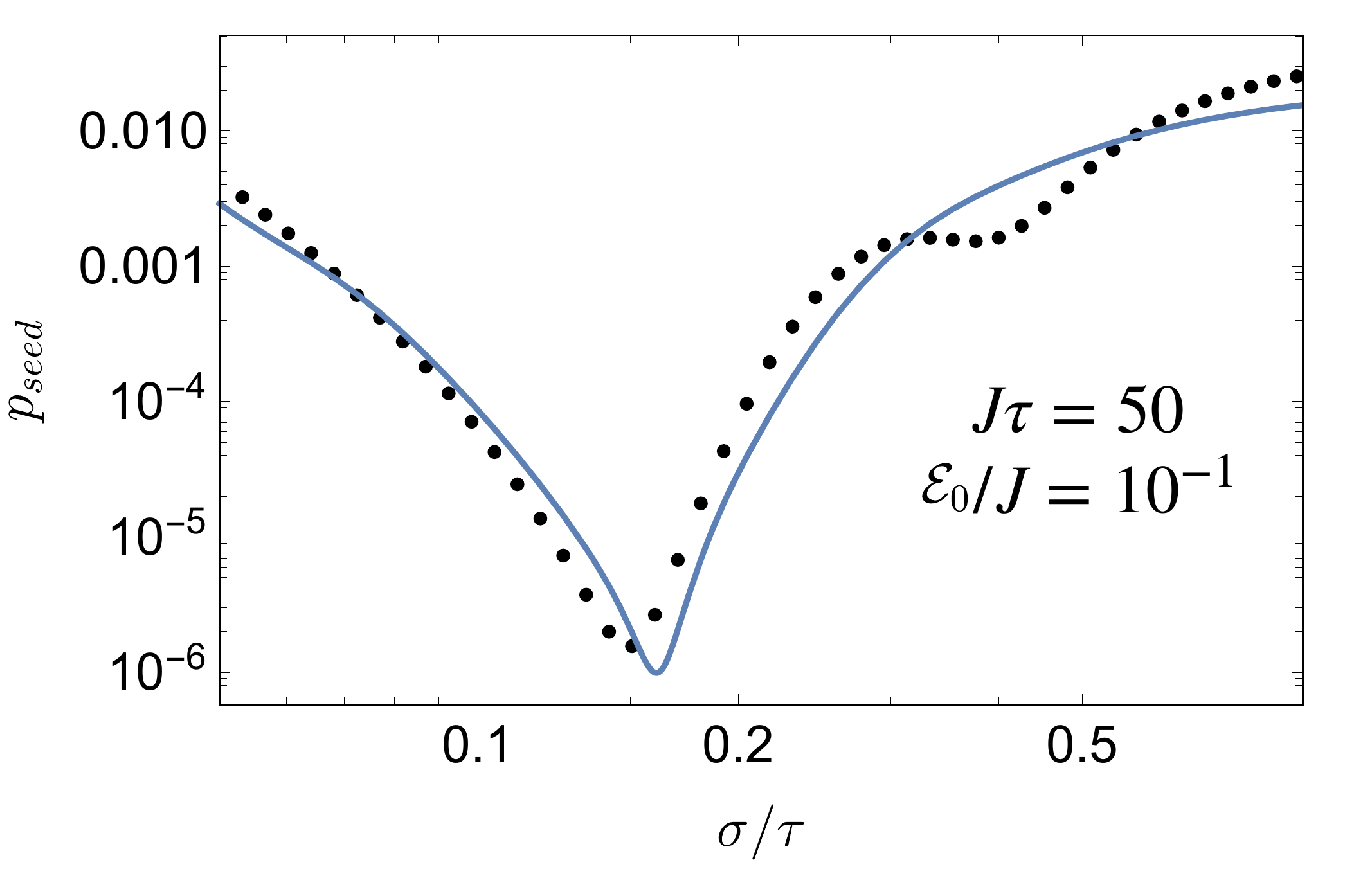}
    \caption{Comparison between the  exact seed creation probabilities computed numerically in a finite size chain with $N=7$ spins (black dots) and the the analytic formula of Eq.~\eqref{eq:p_seed_fin} (blue line); for parameters $J\tau=10$, $\mathcal{E}_0/J=10^{-2}$ (top) and $J\tau=50$, $\mathcal{E}_0/J=10^{-1}$ (bottom).}
    \label{fig:Jtau10}
\end{figure}

We first show that the above estimate, based on first order time-dependent perturbation theory, captures the off-resonant transition probabilities well.
Fig.~\ref{fig:Jtau10}
compares the prediction of Eq.~\eqref{eq:p_seed_fin} with exact numerics done for a chain with $N=7$ spins. The overall agreement is good. The remaining discrepancies are probably due to finite size effects in the numerics, or our neglect of terms of higher order in the non-resonant part of the Hamiltonian, $H_{\rm seed}$. The subset of those processes having an energy mismatch of $J$ only, can contribute with similar amplitude as the one we estimate at first order in $H_{\rm seed}$. 

\FloatBarrier

An explicit expression for the integral in Eq.~(\ref{eq:p_seed_fin}) is not available. However, we will study the seed probability $p_{\rm seed}(\mathcal{E}_0,\sigma,\tau)$ for certain limits,  which are relevant for the optimization problems discussed in the main text.

\begin{figure}[ht!]
    \centering
    \includegraphics[width=0.45\textwidth]{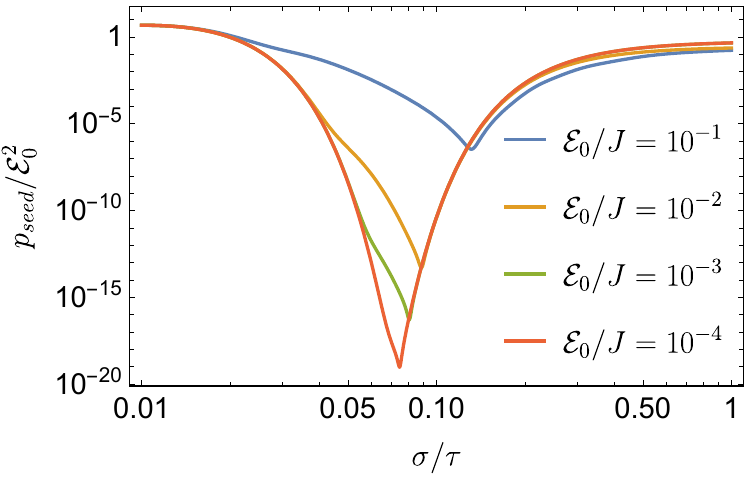}
    \caption{$p_{\rm seed}/\mathcal{E}_0^2$ as a function of $\sigma/\tau$ for $J\tau=100$, evaluated using Eq.~\eqref{eq:p_seed_fin} whereby the series was truncated at $\abs{j}=4$. Note that beyond a sharp minimum, for $\sigma>\sigma_{*}(\mathcal{E}_0)$, there is nearly no dependence on $\mathcal{E}_0$.}
    \label{fig:p_seed}
\end{figure}

Fig.~\ref{fig:p_seed} plots the result of Eq.~(\ref{eq:p_seed_fin}) for $p_{\rm seed}/\mathcal{E}_0^2$ as a function of pulse width $\sigma$ for different values of $\mathcal{E}_0$ (and fixed $\tau$ and $J$). It is seen that the curves collapse beyond a minimum at  $\sigma =\sigma_*(\mathcal{E}_0)$. As explained in the main text, for $\sigma >\sigma_*$ seeds are indeed predominantly created by a single photon process  involving a photon from the tail of the truncated Gaussian, $p_{\rm seed}(\mathcal{E}_0,\sigma>\sigma*,\tau)\approx \frac{\mathcal{E}_0^2}{J^2}e^{-\frac{\tau^2}{4\sigma^2}}$.
We refer to these as tail transitions. When deriving approximations for the formula  Eq.~\eqref{eq:p_seed_fin}, we will instead focus on the regime $\sigma<\sigma_{*}$. 

{We focus on the term $j=1$ in the series~\eqref{eq:p_seed_fin}, as it is representative of the behavior of the full series. We need to estimate
\begin{equation}
    p_{\rm seed}(\mathcal{E}_0,\sigma,\tau)\approx\abs{\int_{-\infty}^{\infty}\mathcal{E}(t)\frac{J_1(4\mathcal{A}(t))}{2\mathcal{A}(t)}e^{iJt}\dd{t}}^2.
    \label{eq:p_seed_inf}
\end{equation}

By Taylor expanding the Bessel function, we obtain the expression
\begin{equation}
    p_{\rm seed}(\mathcal{E}_0,\sigma,\tau)\approx\abs{\sum_{n\geq 0}\int_{-\infty}^{\infty}\mathcal{E}(t)\frac{(-1)^n\left[2\mathcal{A}(t)\right]^{2n}}{n!(n+1)!}e^{iJt}\dd{t}}^2,
    \label{eq:p_seed_2}
\end{equation}
which, by using the Stirling approximation, can be rewritten as the first line of Eq.~(5) of the main text, whereby we drop numerical prefactors $c^n$, with $c=O(1)$, that do not affect our analysis of the leading scaling except for the arguments of logarithms.

We can rewrite this expression as
\begin{equation}
    p_{\rm seed}(\mathcal{E}_0,\sigma,\tau)\approx\abs{\sum_{n\geq 0}\frac{(-1)^n\mathcal{F}\left[\partial_t[4\mathcal{A}]^{2n+1}(t)\right](J)}{2(2n+1)!}}^2,
    \label{eq:p_seed_3}
\end{equation}
which simplifies the calculations below.

This expression applies to any pulse shape. However, to estimate the amplitude for multi-photon processes for $\sigma<\sigma_{*}$, we may neglect the effect of cutting off the pulse, i.e., consider the limit $\tau\to \infty$ (since $\sigma_{*}\ll \tau$). Then, $\mathcal{A}(t)=\sqrt{\frac{\pi}{2}}\mathcal{E}_0\sigma\left[\text{Erf}\left(\frac{t}{\sqrt{2}\sigma}\right)+1\right]$ and, with the following approximation
\begin{equation}
\label{erf}
    \int_{-\infty}^{\infty}dt e^{-i\omega t}\text{Erf}\left(\frac{t}{\sqrt{2}\sigma}\right)\underset{\omega\sigma\gg 1}{\sim} \frac{e^{-\frac{\omega^2\sigma^2}{2}}}{\omega},
\end{equation}
we obtain that $\hat{\mathcal{A}}(\omega)\sim  \frac{\mathcal{E}_0\sigma}{\omega} e^{-\frac{\omega^2\sigma^2}{2}}= \hat{\mathcal{E}}(\omega)/\omega$, up to numerical prefactors and an additional piece $\delta(\omega)$, which turns out to be irrelevant for our analysis at large frequencies.
We can now evaluate the Fourier transform in Eq.~\eqref{eq:p_seed_3} as a $(2n+1)$-fold  convolution of $\hat{\mathcal{A}}(\omega)$,
\begin{eqnarray}
\frac{\mathcal{F}\left[\partial_t\mathcal{A}^{2n+1}(t)\right](\omega)}{(2n+1)!}&= &\frac{\omega}{(2n+1)!} ({\cal A})^{*(2n+1)}(\omega).
\end{eqnarray}

For large frequency $\omega\to J\gg 1/\sigma$ the latter can be evaluated  via saddle point approximation dominated by the frequencies $\omega^* = \pm J/{(2n+1)}$, with a curvature factor $\sim 1/\sigma$ for every frequency integration. This leads to 

\begin{eqnarray}
&&\frac{\mathcal{F}\left[\partial_t\mathcal{A}^{2n+1}(t)\right](J)}{(2n+1)!}
\sim J \frac{1}{\sigma^{2n}} \left[ \frac{\hat{\cal E}(\omega^*)}{\omega^*}\frac{1}{2n+1}\right]^{2n+1}\nonumber \\
&& \quad \sim  J \sigma\left[ \frac{\hat{\cal E}(J/(2n+1))}{\sigma J}\right]^{2n+1}\\
&& \quad \sim \mathcal{E}_0\sigma\left(\frac{\mathcal{E}_0}{J}\right)^{2n} e^{-\frac{J^2\sigma^2}{2(2n+1)}},\nonumber
\end{eqnarray}

whereby we again dropped several factors scaling as $c^n$ with $c=O(1)$. 

With the second but last expression above we can recast Eq.~\eqref{eq:p_seed_3} as  in the second line of Eq.~(5) of the main text.

We finally estimate Eq.~\eqref{eq:p_seed_3} by determining the value $n^*$ which dominates the sum, $n^*(\mathcal{E}_0,\sigma)\sim \frac{J\sigma/2}{\sqrt{2\log{\frac{J}{\mathcal{E}_0}}+O(\log{\log{J\sigma}})}}$. This yields the leading behavior
\begin{equation}
    p_{\rm seed}(\mathcal{E}_0,\sigma,\infty)\sim\exp{-2J\sigma\sqrt{2\log(J/\mathcal{E}_0)}}.
    \label{eq:p_seed_app}
\end{equation}
}
 
\begin{figure}[ht!]
    \centering
    \includegraphics[width=0.48\textwidth]{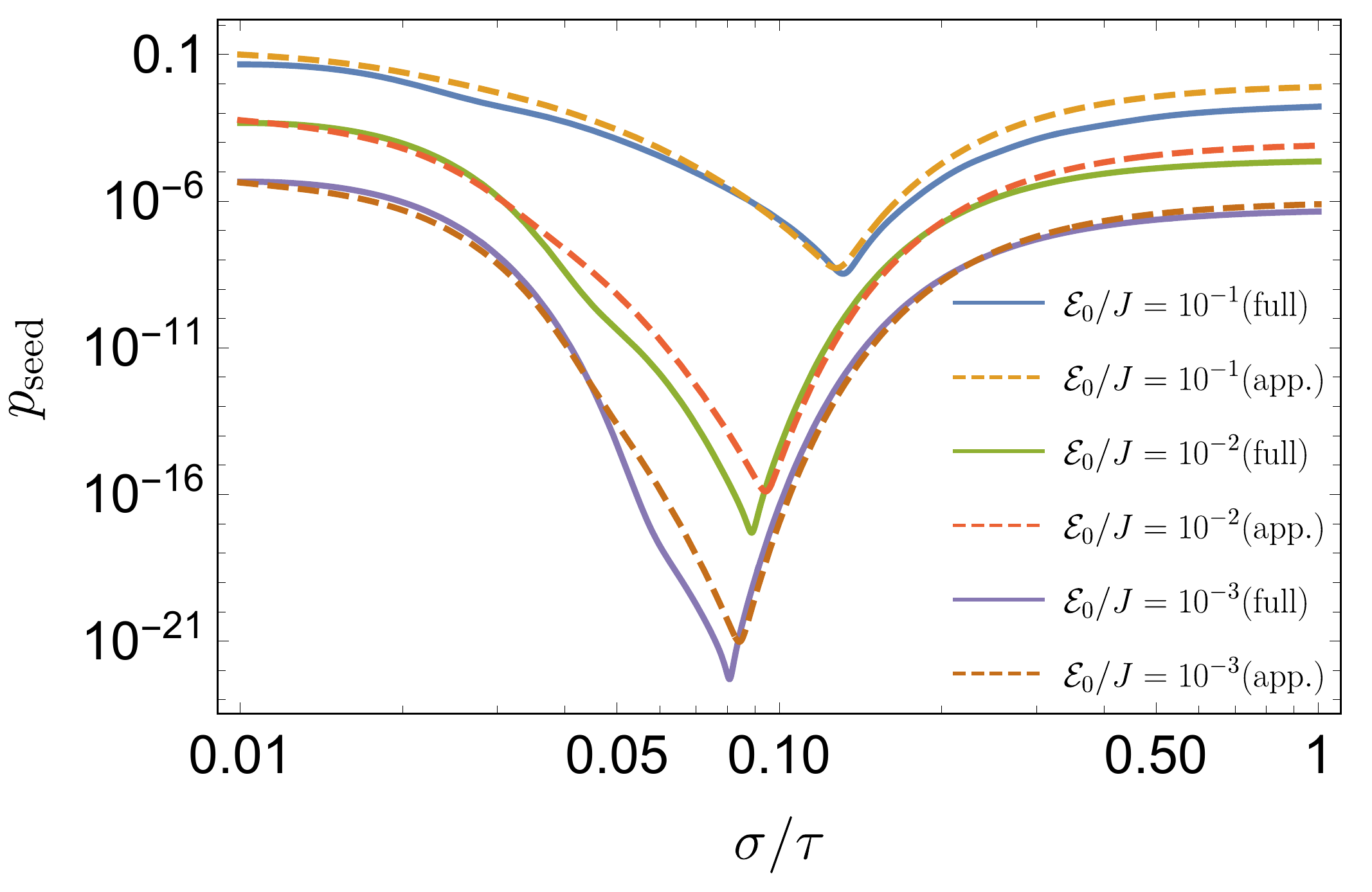}
    \caption{Comparison between $p_{\rm seed}$ as calculated for $J\tau=100$ from Eq.~\eqref{eq:p_seed_fin} (full) and using the approximation of Eq.~\eqref{eq:p_seed_app} (app.) {to which we add the contribution from tail transitions}.}
    \label{fig:fullvsapp}
\end{figure}

In Fig.~\ref{fig:fullvsapp} we show that the obtained approximation Eq.~\eqref{eq:p_seed_app}, {summed with the expression for tail transitions that captures the behavior for $\sigma>\sigma*$} agrees well with the fully evaluated formula of Eq.~\eqref{eq:p_seed_fin}.

\subsection{Hole transitions}
We finally argue that $p_{\rm hole}$ should have the same qualitative dependence on the driving parameters as $p_{\rm seed}$. Indeed, a hole transition corresponds to the formation of two new domain walls (thus, two Majoranas) within the bulk of a cluster. The propagation of these domain walls will  only slightly be modified by the presence of the original domain walls of the cluster if these are distant enough, and thus at the level of the approximations made, nothing significant changes between $p_{\rm seed}$ and $p_{\rm hole}$. 
\subsection{The limit $\tau_R\to \infty$}

We briefly discuss the case in which $\log{J\tau_R}>J\tau_d$, corresponding to exponentially long relaxation times. This implies a lower bound on $\mathcal{E}_0$, called $\mathcal{E}_{0,c}$ in the main text, so small that the condition $\log{\frac{J}{\mathcal{E}_{0,c}}}\ll (J\sigma)^2$ is not satisfied for $\sigma\approx\sigma_{*}(\mathcal{E}_{0,c})$. If $\log{\frac{J}{\mathcal{E}_{0}}}> (J\sigma)^2$, the largest term in the series of Eq.~\eqref{eq:p_seed_3} is the one for $n=0$, and one finds
\begin{equation}
    p_{\rm seed}(\mathcal{E}_0\to 0,\sigma,\infty)\approx 2\pi\mathcal{E}_0^2\sigma^2 e^{-J^2\sigma^2}.
\end{equation}

Since the off-resonant transition is dominated by a single photon event, its probability is minimized by minimizing the spectral density of the pulse envelope at $\omega =J$, which requires $\sigma= \sigma_{*}\approx\sqrt{\frac{\tau_d}{2J}}$.     
The requirement that an initial seed is more likely to grow than to disappear again imposes the lower bound on the drive amplitude $\mathcal{E}_0>\mathcal{E}_{0,c}=\sqrt{\tau_d/\tau_R}/\sigma_* \approx \sqrt{\frac{2J}{\tau_R}}$. 
We conclude that, in this limit
\begin{equation}
    \xi\sim p_{\rm seed}(\sigma_*)^{-1/3}\sim e^{\text{const.}\times J\tau_d}.
\end{equation} 

\section{Phase diagram of the effective RWA Hamiltonian}
In this section we briefly explain the numerical techniques used to establish the quantum phase diagram of the Hamiltonian of Eq.~(8) in the main text.

\begin{figure}[ht!]
    \centering
    \includegraphics[width=0.5\textwidth]{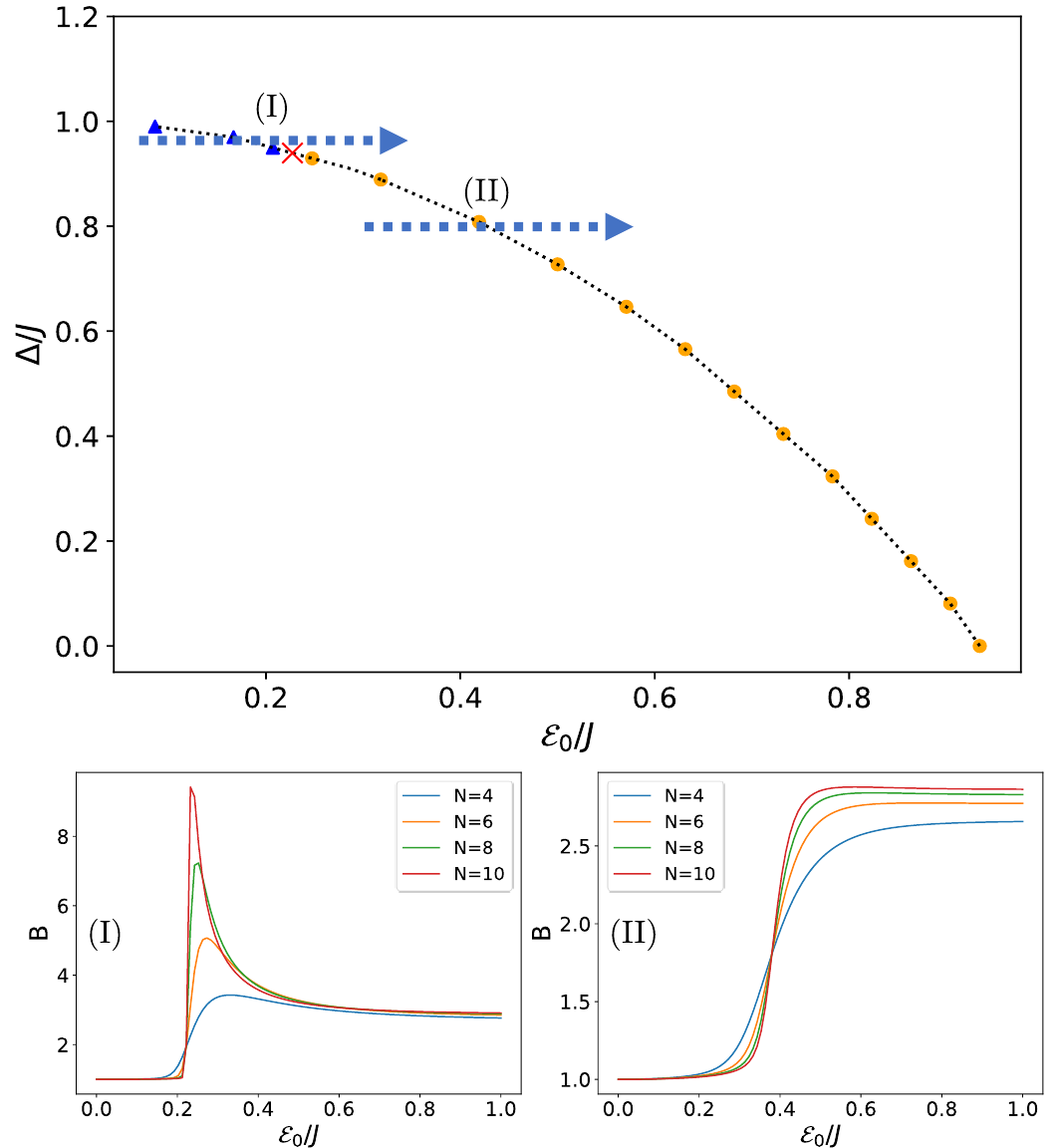}
    \caption{Behavior of the Binder ratio for discontinuous (I) and continuous (II) quantum phase transitions of the Hamiltonian $H_{\text{RWA}}$ of Eq.~(8) in the main text.}
    \label{fig:binder}
\end{figure}

A common numerical technique used to distinguish continuous from discontinuous phase transitions is the analysis of the Binder ratio \cite{binder1981static}. This statistical quantity is defined as the ratio $B=\frac{\ev{m^4}}{{\ev{m^2}}^2}$ where $m$ is the order parameter, in our case (for a chain of size $N$) $m=\frac{1}{N}\sum_{j=1}^{N}(-1)^j S_j^z$. Deep in the ordered phase, the ratio of the numerator and the denominator of $B$ is equal to 1. In the disordered phase instead, the distribution of the order parameter is a Gaussian centered at 0. Hence, $B=1$ in the ordered and $B=3$ in the disordered phase, respectively. For second order transitions in finite size systems, this step-like jump is smoothed into a continuous evolution from one limit to the other. The critical point is identified via finite size scaling, as the crossing point of curves showing the Binder ratio for different system sizes. For first order transitions, instead, a phase coexistence window of extent $\sim 1/N$ around criticality exists in the disordered phase. This causes a volumetric ($\sim N$) divergence of the Binder ratio \cite{PhysRevB.96.115160,Iino2019DetectingSO}. An accurate analysis of the finite size scaling allows to determine both the phase boundaries and the nature of the quantum phase transition of the Hamiltonian $H_{\text{RWA}}$ of Eq.~(8) in the main text. As an illustration, Fig.~\ref{fig:binder}, shows numerical results for a discontinuous and a continuous transition. 

\bibliography{supp}